# Persistent quantum interfering electron trajectories

J. E. Kruse[1], P. Tzallas[1*], E. Skantzakis[1] and D. Charalambidis[1,2]

[1]*Foundation for Research and Technology—Hellas, Institute of Electronic Structure and Laser, PO Box 1527, GR711 10 Heraklion, Crete, Greece*

[2]*Department of Physics, University of Crete, PO Box 2208, GR71003 Heraklion, Crete, Greece*

*Corresponding author e-mail address: ptzallas@iesl.forth.gr

The emission of above-ionization-threshold harmonics results from the recombination of two electron wavepackets moving along a "short" and a "long" trajectory in the atomic continuum. Attosecond pulse train generation has so far been attributed to the short trajectory, attempted to be isolated through targeted trajectory-selective phase matching conditions. Here, we provide experimental evidence for the contribution of both trajectories to the harmonic emission, even under phase matching conditions unfavorable for the long trajectory. This is finger printed in the interference modulation of the harmonic yield as a function of the driving laser intensity. The effect is also observable in the sidebands yield resulting from the frequency mixing of the harmonics and the driving laser field, an effect with consequences in cross-correlation pulse metrology approaches.

Harmonic generation above the ionization threshold is governed by the recombination of localized electron wave packets ejected into the continuum and driven back towards the core upon reversal of the linearly polarized driving field [1]. According to this model, at a given driving laser intensity $I_L$, two interfering electron trajectories with two different total flight times $\tau_q^L(I_L)$ and $\tau_q^S(I_L)$ contribute to the emission of each harmonic $q$. $L$ and $S$ stand for "long" and "short" [1]. The electron wave packet phases, accumulated during their motion in the continuum, and with it the phases of the emitted harmonics are approximated by $\varphi_q^{L,S}(I_L) \approx -U_p \tau_q^{L,S}(I_L)$ (where $U_p$ is the ponderomotive energy) and are proportional to the product $\tau_q^{L,S}(I_L) \cdot I_L$ [1]. The driving-intensity-dependent phase difference of the two trajectories $\Delta\varphi_q^{L,S}(I_L) = \varphi_q^L(I_L) - \varphi_q^S(I_L) \propto \left[\tau_q^L(I_L) - \tau_q^S(I_L)\right] \cdot I_L \neq 0$, leads to alternating constructive and destructive interference. This results in a modulation of the harmonic yield $Y_q(I_L) \propto \left| E_q^S \, \mathrm{e}^{-i(\omega_q t + \varphi_q^S(I_L))} + E_q^L \, \mathrm{e}^{-i(\omega_q t + \varphi_q^L(I_L))} \right|^2$. As the driving laser intensity $I_L$ increases, the difference in flight times $\Delta\tau_q^{L,S}(I_L) = \tau_q^L(I_L) - \tau_q^S(I_L)$ increases. This results in a reduction of the period of the harmonic yield's modulation. On the other hand, for higher harmonic orders $q$, the difference in flight time $\Delta\tau_q^{L,S}(I_L)$ is shorter, resulting in a slower modulation of the harmonic yield. Figure 1(a) shows the dependence of $\Delta\varphi_q^{L,S}$ (for $q$=13[th] to 17[th]) on the intensity of the driving laser field, calculated by solving of the quantum mechanical three step model [1]. For plateau harmonics the dependence of $\Delta\varphi_q^{L,S}$ is close to linear, while when the harmonics are close the cut-off region (in the present case the 17[th] harmonic) it becomes non-linear. $\Delta\varphi_q^{L,S}$ stays close to zero, for above cut-off harmonics. This behavior can be seen in Fig. 1(b) which shows the dependence of $d(\Delta\varphi_q^{L,S}(I_L))/dI_L$ on the driving laser intensity. The so far discussed atomic response leads to emission of harmonic radiation that is further modified through propagation, which affects the emitted yield through phase matching. The condition for optimal phase matching of focused Gaussian beams is $\vec{k}_q = q\vec{k} + \Delta\vec{k}_g + \Delta\vec{k}_d + \vec{\nabla}\varphi_q^{L,S}$ with $\vec{k}_q$ and $\vec{k}$ being the k-vectors of the $q^{th}$ harmonic and the fundamental, respectively, $\Delta\vec{k}_g$ the Gouy phase and $\Delta\vec{k}_d$ the mismatch caused through dispersion [1]. This condition favors far off-axis generation by the long trajectory and on axis generation by the short trajectory only if the laser focus is before the jet [2, 3]. For this reason, the interference effect has been mostly studied in detail for off-axis harmonic generation [4]. In the present work we demonstrate on-axis interference of the two trajectories for all three focus positions, namely the focus before, on and after the gas jet. These are also the conditions under which the effect was so far not expected. The observed interference manifests the contribution of both trajectories in all three geometries, even though with different weighting factors [5]. The interference effect has recently been observed on axis also for *below-threshold* harmonic generation at focus [6].

The experiments have been conducted using the FORTH-ULF 4TW Ti:Sapphire laser system delivering pulses of 40 fs duration at 10 Hz, central wavelength 800 nm and energy up to 150 mJ/pulse. The experimental set-up used in the present study is described in detail in ref. [5, 7]. An annular laser beam was focused by a 3 m focal length lens into a xenon pulsed gas jet, where the odd harmonics were generated. A mask with a hole in the center was placed in the laser beam to create a 1.5 mm diameter IR beam (dressing-beam) in the center of the annular beam. The temporal delay between the IR-dressing-beam and annular beam was zero. After the xenon jet a silicon wafer was placed at the fundamental's Brewster angle of 75° to reduce the IR radiation and to reflect the harmonics [8] towards the detection area. A λ/2 waveplate was used to rotate the polarization of the laser by a very small angle (< 3°) in order for a small fraction of the central IR dressing-beam to be reflected by the wafer towards the detection area. The reflected two-color beam was passing through a 2 mm diameter aperture, placed 2 m downstream the Xe jet, which was blocking further any residual part of the annular IR radiation and was selecting only the central part of the beam cross section. Then the beam was focused by a gold coated spherical mirror of 5 cm focal length into an argon pulsed gas jet. For the measurement of the interference effect between the two electron trajectories, photoelectron (PE) spectra of Argon ionized in the presence of the XUV and IR field have been recorded as a function of the laser field intensity. The PE spectra consist of a series of 11$^{th}$-17$^{th}$ harmonic single-photon ionization peaks and additional two-photon ionization (IR+XUV) "sideband" peaks S12-S16 appearing between them. The intensity of the laser at the harmonic generation region was varied between ~ $10^{14}$ and ~ $4 \times 10^{13}$ W/cm$^2$. Measurements have been performed for three different positions of the laser focus with respect to the position of the gas jet, I) $z_{BJ} = -0.43b$ (laser focus before jet), II) $z_{OJ} = 0$ (laser focus at jet) and III) $z_{AJ} = +0.28b$ (laser focus after jet), respectively. $b$=18 cm is the confocal parameter, assuming Gaussian beam geometry. It is known from modeling including propagation effects in the generating medium [2] that phase matching favors in case I) the contribution of the "short" trajectory, in case II) both and in case III) the contribution of the "long" trajectory to the harmonic emission. In our case, the contribution of the long trajectory has been further reduced by the on axis spatial filtering mentioned above.

Fig. 2 shows the measured 13$^{th}$ and 15$^{th}$ harmonic and the S12-S16 sideband yield in dependence on the laser intensity for the three positions of the laser focus.

The upper panel shows an example of raw data points (grey dots) of harmonic 13 together with a 10 points moving average over the raw data (red line). A low contrast modulation is observable superimposed on a "background" signal, which increases nonlinearly with the laser intensity. The middle and lower panels show the measured modulation for harmonics 13[th] and 15[th] and sidebands S12-S16 after subtraction of the background signal for all three geometries. In all cases the signals present the following characteristic features:

a) They feature a periodic modulation, with double peak interference maxima. This is more pronounced when focusing before the jet and at the jet. The average period of the oscillations is $I/I_{max}$=0.23, which results in an oscillation period of $\approx$ 0.23x10$^{14}$ W/cm$^2$ for $I_{max} \approx 10^{14}$ W/cm$^2$. This is in reasonable agreement with the expected value [1].

b) The distance between the maxima and/or minima (green lines in Fig. 2) when focusing before and at the jet becomes smaller for higher laser intensities, in agreement with the theoretical predictions given in Fig.1. When focusing after the jet this behavior is not observed. This could be due to the not observed double minima structure, leading to an uncertainty whether an observed minimum is a "main" or a "secondary" one.

c) The fringe contrast, which depends on the percentage of the "long" trajectory contribution, becomes higher as the focusing geometry changes from "before jet" to "after jet". The harmonic and sideband fringe contrast, when focusing before, at and after the jet, is $\approx$ 9 %, $\approx$ 12% and $\approx$ 25%, respectively.

d) The position of the interference minima shifts from one sideband to the next. This shift, which is shown by the tilted green lines in lower panels of fig. 2, depends on the intensity of driving laser field, the sideband order $(q+1)$ and the focusing geometry.

When focusing before or at the jet the interference minima are shifted to lower laser intensities for the higher sideband orders. The effect is stronger, when the focus is before the jet. When the focus is after the jet, no systematic shifts have been observed. This could relate to the sensitivity of the long trajectories on the variations of the driving laser intensity.

e) Interestingly, even the harmonics close to the cut-off region (like $17^{th}$) show an interference modulation. This observation, shown in the modulation signal of the S16 in Fig. 3, is compatible with the results of the quantum mechanical three step model, according to which both trajectories are contributing to the cut off harmonics, but the difference [$\tau_{\text{cut-off L}}$-$\tau_{\text{cut-off S}}$] is much smaller than for plateau harmonics. This is manifested in the observed long modulation intervals.

Considering both the short and long trajectories contributing to the harmonic generation process, the modulation of the sideband signal ($S_{q+1}$) (the index $(q+1)^{th}$ indicates the order of the sideband) with the laser intensity and delay $\tau$ reads

$$S_{q+1}(\tau, I_L) \propto \cos(2\omega_L \tau + \Delta\varphi^S_{q,q+2}) + \cos(2\omega_L \tau + \Delta\varphi^L_{q,q+2}) +$$

$$+ 2\cos(A_{q+1})\cos(2\omega_L \tau + \frac{\Delta\varphi^S_{q,q+2} + \Delta\varphi^L_{q,q+2}}{2}) + \cos(\Delta\varphi^{SL}_q) + \cos(\Delta\varphi^{SL}_{q+2}) \qquad (1)$$

,where $A_{q+1} = \frac{1}{2}(\Delta\varphi^{SL}_q + \Delta\varphi^{SL}_{q+2})$, $\Delta\varphi^S_{q,q+2} = \varphi^S_q - \varphi^S_{q+2}$, $\Delta\varphi^L_{q,q+2} = \varphi^L_q - \varphi^L_{q+2}$, $\Delta\varphi^{SL}_q = \varphi^S_q - \varphi^L_q$, $\Delta\varphi^{SL}_{q+2} = \varphi^S_{q+2} - \varphi^L_{q+2}$. Here, $\varphi^S_q, \varphi^L_q$ are the phases of harmonics generated by electrons from the short and long trajectory, respectively, and $\omega_L$ is the fundamental laser frequency. In eq. 1 the atomic phase shift is assumed to be zero due to it's negligibly influence on the sideband signal. Further it is assumed that the harmonic intensity does not depend on the laser intensity and $I_q^S = I_q^L = I_{q+2}^S = I_{q+2}^L$. $I_{q,q+2}^S$ and $I_{q,q+2}^L$ are the normalized intensities of the harmonic order $q$ and $q+2$ generated by the short and long trajectories, respectively. The phases of the harmonics were calculated from the quantum mechanical version of the three step model [1]. Propagation effects in the harmonic generation medium were not taken into account. Figure 3a,b,c shows the modulation of the sideband signal S12-S16 as a function of the delay $\tau$ and the laser intensity ($I_L$). The dashed lines depict the intensity values where the harmonics contributing to the sideband $q+1$, reach the cut-off. A line-out of Fig's. 3a,b,c at $\tau$=2.67 fs ($2\pi$ delay between XUV and IR) shows the dependence of the sideband signal as a function of $I_L$ at the experimental conditions of the present work (Fig. 3d). The blue lines in Fig. 3d connect the position of the interference minima between the sidebands. All characteristic features (a)-(e) of the experimental results are in reasonable agreement with the calculated ones. A periodic modulation (with average period of $0.2 \times 10^{14}$ W/cm$^2$) with double peak interference maxima has been calculated. The distance between the maxima and/or minima becomes smaller at higher laser intensities. The fringe contrast (not shown) depends on the percentage of the "long" trajectory contribution. The relative position of the interference minima and/or maxima between the different sidebands (blue lines) depends on the intensity of the driving laser field and the sideband order ($q+1$). The harmonics close to the cut-off region show an interference modulation. The shift to lower laser intensities of the sideband interference minima, which is found to be more pronounced in the calculated than in the experimental trace, could be attributed to propagation effects that are not taken into account in the present model. The intensity dependence of the harmonic phase further results in an intensity dependence of the pulse duration in the attosecond pulse train. Due to the stronger intensity dependence of the long trajectory phase, a driving intensity modulation may lead to destruction of the attosecond confinement, when both trajectories contribute to the XUV emission. On axis harmonic generation by laser beams focused before the gas jet was so far considered to eliminate the long trajectory. The present findings, in agreement with those of our relevant recent work [5] on a comparative study between the $2^{nd}$ order intensity volume autocorrelation (IVAC) [9] and the RABITT [10] technique, do not sustain this assumption, evidencing implications to the emitted pulse durations and their metrology.

Concluding we show that both the long and short trajectories are contributing to "on axis" harmonic generation in a gas medium. We show this for both plateau and cut-off harmonics, at different phase matching conditions. Phase matching affects to a certain extent the relative contribution of the two trajectories, but essentially it does not eliminate any of the two. The results of this work are in full agreement with pulse duration measurements of attosecond pulse trains through $2^{nd}$ order IVAC [9], but in conflict with those conducted with the RABITT approach [10]. Thus, they play a significant role for the accuracy of measuring attosecond pulses by means of cross-correlation techniques. They further contribute to the improvement of the accuracy of atomic-molecular tomography techniques [11] and precision measurements by extending the frequency combs into the XUV spectral region [6, 12].

**Acknowledgements**

This work is supported in part by the Ultraviolet Laser Facility (ULF) operating at FORTH-IESL (contract HPRI-CT-2001-00139), the ELI research infrastructure preparatory phase program, the FASTQUAST ITN.

**Figure captions**

**Figure 1.** (a) Dependence of $\Delta\varphi_q^{L,S}$ (for $q$=13$^{th}$ to 17$^{th}$) on the intensity of the driving laser field, calculated by solving of the quantum mechanical three step model. (b) Dependence of $d(\Delta\varphi_q^{L,S}(I_L))/dI_L$ on the intensity of the driving laser field..

**Figure 2.** Measured dependence of the 13$^{th}$, 15$^{th}$ harmonic and S12-S16 sideband yield on the laser intensity for the three positions of the laser focus. In the upper panel is shown an arbitral example of raw data points (grey dots) of harmonic 13$^{th}$ together with a 10 points moving average of the raw data (red line). In the central and lower panel the measured modulation for harmonic 13$^{th}$, 15$^{th}$ and S12-S16, respectively are shown after subtraction of the "background" signal for all three geometries.

**Figure 3.** (a), (b), (c) Modulation of the sideband signal S12-S16 as a function of the delay $\tau$ and the laser intensity ($I_L$). (d) A line-out of Figs. 3a,b,c at $\tau$=2.67 fs (equal to zero delay between XUV and IR) shows the sideband signal as a function of $I_L$.

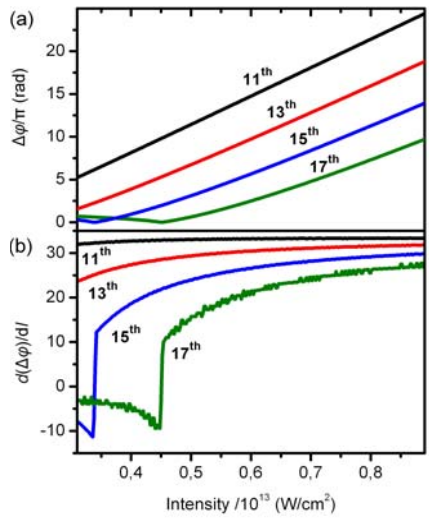

**Figure 1.**

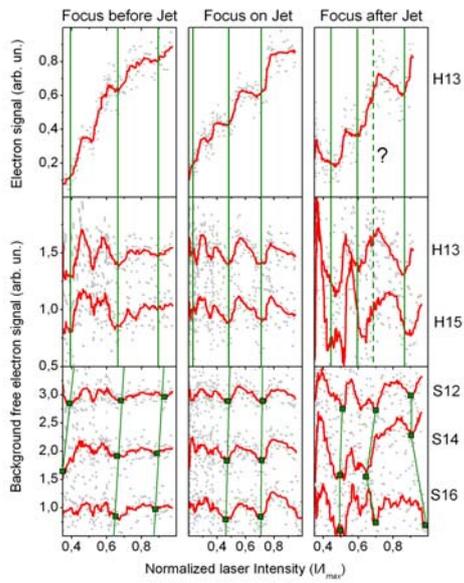

**Figure 2.**

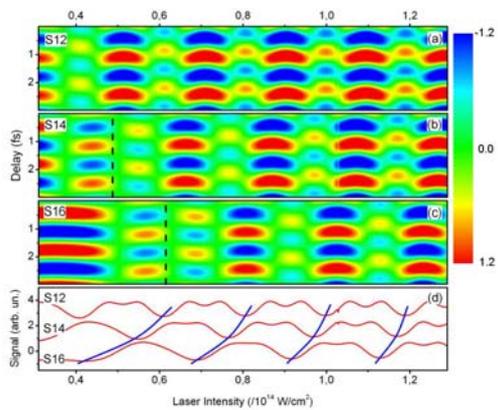

**Figure 3.**